# Ultrafast time-resolved Faraday rotation in EuO thin films

F. Liu[1,2], T. Makino[3*], T. Yamazaki[4], K. Ueno[5,6], A. Tsukazaki[1,6], T. Fukumura[7,6], Y. Kong[2], and M. Kawasaki[1,3,8,9]

[1]Quantum Phase Electronics Center and Department of Applied Physics, University of Tokyo, Tokyo 113-8656, Japan,

[2]School of Physics, Nankai University, Tianjin 300071, China,

[3]Cross-correlated Materials Research Grooup (CMRG) and Correlated Electron Research Group (CERG), RIKEN Advanced Science Institute, Wako 351-0198, Japan

[4]Institute for Materials Research, Tohoku University, Sendai, 980-8577, Japan,

[5]Graduate School of Arts and Sciences, University of Tokyo, Tokyo 153-8902, Japan

[6]PRESTO, Japan Science and Technology Agency, Tokyo 102-0075, Japan,

[7]Department of Chemistry, University of Tokyo, Tokyo 113-0033, Japan,

[8]WPI-Advanced Institute for Materials Research, Tohoku University, Sendai 980-8577, Japan,

[9]Japan Science and Technology Agency (CREST), Tokyo 102-0075, Japan

**Abstract**

We have investigated the ultrafast spin dynamics in EuO thin films by time-resolved Faraday rotation spectroscopy. The photoinduced magnetization is found to be increased in a transient manner, accompanied with subsequent demagnetization. The dynamical magnetization enhancement showed a maximum slightly below the Curie temperature with prolonged tails toward both lower and higher temperatures and dominates the demagnetization counterpart at 55 K. The magnetization enhancement component decays in ~1 ns. The realization of the transient collective ordering is attributable to the enhancement of the *f-d* exchange interaction.

PACS numbers: 78.20.Ls, 42.50.Md, 78.30.Hv, 75.78.Jp

Ultrafast magneto-optical experiments in ferromagnets have attracted considerable interests in condensed-matter physics for understanding the dynamics of electrons's spin degree of freedom [1]. Most experiments to date showed an ultrafast demagnetization due to laser-induced electronic heating [2]. There is stronger interest in ultrafast photo-enhancement of magnetization. In spite of the recent observations of the ultrafast magnetization enhancement in strongly correlated manganites [3,4] and a diluted magnetic semiconductor[5], the understanding of the mechanism for this nonequilibrium phenomenon [1] requires accumulation of theoretical efforts and experimental results in a variety of ferromagnetic systems. In this study, we investigate the spin dynamics in EuO thin films, which are expected to be suitable for the observation of the photoinduced magnetization enhancement. EuO is a ferromagnetic semiconductor with a band gap of about 1.2 eV [6,7] and the conduction-band spin split by 0.6 eV [8,9]. The magnetic properties of pure EuO have been explained by Heisenberg model because its magnetic moment of 6.9 $\mu_B$/Eu is similar to that of a free $Eu^{2+}$ ion [6,7]. Its Curie temperature ($T_C$), for thin films, increases with electron doping from the intrinsic value of around 69 K [10,11], up to 125 K reported for 4% Gd-doped [6,12,13] or by oxygen deficient samples [14,15,16]. It is known that the increase in the Curie temperature and the insulator-to-metal transition have been explained in terms of the effect of magnetic coupling, *i.e.*, the concept of the magnetic polaron, where 5*d* electrons interact with the local 4*f* magnetic moments [7]. An observation of the transient photo-induced enhancement was reported in EuO in the magnetic field (*B*) up to *B* = 0.2 T [17]. It is desirable to apply the statruation magnetic field ($B \approx 3$ T) for an unambiguous discussion. Above-gap photoexcitation has been known to inject the $5d$ electrons and expected to create the magnetic polaron in EuO [7].

---

[*] e-mail: tmakino@riken.jp

In this paper, we study the resonant effect of the photo-induced spin and charge dynamics in EuO and report the observation of the ultrafast photoinduced enhancement of magnetization. The data show magnetization enhancement, followed by subsequent demagnetization. The temperature profile of demagnetization is monotonically decreasing up to $T_C$, while the transient magnetization enhancement has a maximum slightly below $T_C$ and persists well above $T_C$.

EuO thin films were grown by a pulsed-laser deposition method with using a Eu metal as a target on YAlO$_3$ (110) single crystalline substrates. The substrates were annealed at 1200 °C in a furnace in air prior to the deposition to form an atomically flat surface. The base pressure of the PLD chamber was $8 \times 10^{-10}$ Torr, while the oxygen partial pressure ($p_{O2}$) during the growth was set at $1 \times 10^{-7}$ Torr. The EuO films were then capped with AlO$_x$ films *in-situ* to avoid degradation of the films in air. EuO and AlO$_x$ layers have thicknesses of 310 and 30 nm. The film turned out to be too insulating to be quantified by a conventional transport measurement method. The magnetization curves were measured with a superconducting quantum interference device magnetometer. Details are described in Ref. [15]. For the time-resolved Faraday rotation ($\Delta\theta_F$) and pump-probe transmission ($\Delta T/T$) measurements, a Ti:sapphire regenerative amplifier system (1.55 eV, 100 fs, and 1 KHz) was used to excite an optical parametric amplifier. Time resolution of the measurement system is ~200 fs. The pump fluence at the sample surface was approximately 0.5 mJ/cm$^2$. The magnetic field applied normal to the sample surface is $B$ = 3.2 T. For the steady-state Faraday rotation measurements, a continuous-wave semiconductor laser diode is used.

Figure 1 shows the basic magnetic [(a) and (b)] and optical [(c) and (d)] properties of EuO thin films. Magnetization is displayed as a function of temperature under in-plane and out-of-plane configurations. This surely amounted to lowest-temperature saturation magnetization of 6.9 $\mu_B$/Eu, which is in reasonable agreement with the available literature data [6,7]. The inverse plot of the magnetic moment yielded in a Curie temperature ($T_C$) of 70 K, close to the previously reported value of 69 K. The calculated Brillouin $J = 7/2$ function result (solid line) in Fig. 1(a) is in a good agreement with the experimental data. The in-plane and out-of-plane magnetic field dependences of magnetization are plotted in Fig. 1(b) at 5 K. The sample showed out-of-plane saturation fields of $|B_{sat}| \approx 3.2$ T [13]. Figure 1(c) shows an optical absorption spectrum of a 50-nm-thick thin film taken at 300 K. Absorption bands which can be seen which can be attributed to the transitions schematically shown in Fig. 1(d) [7]. We have chosen the excitation photon energy (1.91 eV) to promote the $4f \rightarrow 5dt_{2g}$ transition.

We show the time trace of $\Delta T/T$ at 10 K in Figs. 2(a) and 2(b) to see the charge dynamics. The time profile consists of two components: the ultrafast component whose decay time constant is ~10 ps and slower component that decays in several nanoseconds. The first and second components are attributable to the cooling process of the hot electrons and the decay of the electron-hole pairs probably trapped in metastable states, respectively. Figures 2(c) and 2(d) represent time evolution of $\Delta\theta_F$. The lineshape is cantly different from that of $\Delta T/T$. We confirmed that the reversal of the applied magnetic field direction changed the polarity of the signal [18] The coincidence between pump-induced ellipticity and rotation [5,19] was also confirmed. The slightly-detuned excitation at 1.55 eV yielded in much weaker pump-induced rotation. These are supporting evidences that one observes the resonance effect of magnetization dynamics in this sample. The time trace includes two dynamic magnetization processes; one is an enhancement of magnetization ($\Delta\theta_F > 0$) having two decay components, the other is a subsequent demagnetization at larger time delays ($\Delta\theta_F < 0$).

Figure 3 shows $\Delta\theta_F$ for different measurement temperatures on short (a) and long (b) time scales. The magnetization [red (medium gay) shaded region) and demagnetizations [blue (dark gray) shaded region] show different temperature dependences on each other. The photoenhanced magnetization is seen over the whole temperature range, $T \leq 250$ K. This is also seen in the traces monitored at 2 and 26 ps in Fig. 3(c). The trace at 2 (26) ps delay reflects the temperature dependence of faster (slower) decaying component. On the other hand, the slow demagnetization is seen at larger delay time (860 ps), which vanishes for $T > 55$ K. Unlike the case of the transition-metal ferromagnets, such a slow demagnetization is rather reasonable in a system containing 4$f$ electrons [20]. The photoexcitation mainly raises the effective temperature of the 5$d$-band electrons instantaneously. If the magnetic moments are contributed from these 5$d$ electrons, the fast demagnetization can be observed through the spin-flip scattering occurring only in the 5$d$ band, which is not the case in this material. The effective temperature of the 4$f$ spins determining the demagnetization then approaches to that of the heated 5$d$ spins gradually through equilibration between two spin baths. This leads to the slow demagnetization observed. As shown in Figure 3(c), at elevated lattice temperatures, the demagnetization components become obscure and overwhelmed by the enhancement counterpart.

Having established the observation of photoinduced enhancement of magnetization in EuO, we decompose the photoinduced Faraday rotation $\Delta\theta_F$ into magnetization enhancement ($\Delta\theta^{en}_F$) and demagnetization ($\Delta\theta^{de}_F$) components based on their different time scales. Figure 4(a) shows a time-trace of $\Delta\theta_F$ taken at $T = 10$ K and at $B = 3.2$ T, which is described by an equation $\Delta\theta^{en}_{F1} \exp(-t/\tau_{en1}) + \Delta\theta^{en}_{F2} \exp(-t/\tau_{en2}) + \Delta\theta^{de}_F [1 - \exp(-t/\tau_{de})]$. Here, the $\Delta\theta^{en}_{F1(2)}$ and $\tau_{en1(2)}$ in the first (second) term are the magnitude and decay time constant of magnetization enhancement [red (medium gray) line)], while the quantities related to demagnetization [blue ( dark gray) line) are represented by $\Delta\theta^{de}_F$ and $\tau_{de}$. The decay or buildup time constants are $\tau_{en1} = 5$ ps, $\tau_{en2} = 1.4$ ns, and $\tau_{de} = 76$ ps. $\tau_{en2}$ is related to the diffusion or recombination time of the magnetic polarons, whereas $\tau_{de}$ is caused by the indirect heat transfer from the 5$d$ system to the 4$f$ system as has been mentioned already.

Figures 4(b)-4(c) plots the preexponential terms of $\Delta\theta^{en}_{F2}$, and $-\Delta\theta^{de}_F$ against temperature by closed symbols. The demagnetization $-\Delta\theta^{de}_F$ profile [Fig. 4(c)] is a monotonically decreasing function of temperature and similar to that of the steady-state Faraday rotation $\theta_F$ for the same sample evaluated [inset of Fig. 4(c)]. The experimental results [red (medium gray) line)] are in reasonably good agreement with the Brillouin $J = 7/2$ function (solid black line). On the other hand, the magnetization enhancement $\Delta\theta^{en}_{F2}$ apparently has a different temperature profile peaking slightly below $T_C$ shown by downward arrows in Fig. 4(b). The amplitude of peak for $\Delta\theta^{en}_{F2}$ amounts to $\approx 0.04\%$ of the static $\theta_F$ (4.6 degrees at 10 K). The observed magnetization enhancement even at low temperatures is reminiscent to the dynamical magnetization enhancement induced by the magnetic polarons in a diluted magnetic semiconductor [21,22].

We first discuss the peaking behavior slightly below $T_C$, related to the reorganization effect of the partially-randomized 4$f$ magnetic moments within the Bohr radius of the magnetic polarons [21]. This should manifest itself as an increase in the $T_C$ ($\Delta T_C$), which has been demonstrated in the electron-doped EuO. Wang *et al.* have attributed such a kind of peaking behavior in GaAs:Mn to the increase in $T_C$ ($\Delta T_C$) [5,22]. We performed least-square fit to the experimental data using a model based on the effect of $\Delta T_C$ on the differential magnetization ($\Delta M$), which is proportional to $\Delta\theta^{en}_{F2}$ in the framework of Weiss mean-field theory [23]. We calculate the differential magnetization $\Delta M$ based on the effect of $\Delta T_C$. The experimental $M$ and $\theta_F$ curves [inset of Fig. 3(c)] are calculated [23]:

$$M(T,\Delta T_C) = \mu N B_J \left( \frac{\mu}{k_B T} + \frac{3J(T_C+\Delta T_C)M(T,\Delta T_C)}{(J+1)N\mu^2} \right) \quad (1)$$

where $B_J(x)$ is the Brillouin function, $N$ is the density, and $k_B$ is Boltzmann constant, and $T$ is a lattice temperature. The Eu total angular momentum $J$ is 7/2. At each $\Delta T_C$, the magnetization enhancement $\Delta M$ was solved self-consistently as a function of $T$. Bearing a relationship of $\Delta\theta^{en}_{F2}/\theta_F = \Delta M/M$ in mind, the calculated $\Delta M\theta_F/M$ (a dashed curve) and experimental $\Delta\theta^{en}_{F2}$ (solid symbols) are shown in Fig. 4(b) for $\Delta T_C = 30$ mK. Note, the solid curves were obtained by the multiplication between the calculated $\Delta M/M$ and the experimental $\theta_F$. The calculated $\Delta\theta^{en}_F$ curve has an asymmetrical line shape. The prolonged tail toward $T > T_C$ somehow reproduces the experimental $\Delta\theta^{en}_{F2}$ values in the high temperature range. On the other hand, the calculated curve reduces asymptotically to zero when the temperature decreases because the saturation magnetization remains unchanged at low temperatures, which is not the case in the experimental data.

Bearing in mind that the magnetic polaron contains a 5$d$ electron as a constituent and it can polarize the 5$d$ spin efficiently at low temperatures, the effect of the spin polarization of the 5$d$ electrons can explain the observed magnetization enhancement at low temperatures [21]. Barbagallo *et al.* attributed the increase in the magnetic moment even at 5 K in oxygen-deficient EuO to the preferential population of the 5$d$ electrons in the majority spin branch of the spin-polarized conduction band [14], which may have the same origin with our observation in the low temperature range. It is reasonable that the 5$d$ spin polarization degree, which should have monotonically decreasing function of the temperature [7], affects the $\Delta\theta_F$ curve in the low temperature range. The above-mentioned two contributions, which originated from the duality in the role of the magnetic polaron, could explain the overall temperature dependence as shown in Fig. 4(b).

In summary, we have demonstrated ultrafast photoenhanced ferromagnetism in EuO. Our data clearly show that the dynamic magnetization buildup occurs, accompanied with subsequent demagnetization in time-resolved Faraday rotation trace. An argument based on the spin-polarization of the photoinjected 5$d$ electrons and reorganization of 4$f$ local moments is successful in explaining the overall temperature dependence of our observed magnetization enhancement.

Acknowledgements—the authors thank K. Katayama, M. Ichimiya, and Y. Takagi for helpful discussion. This research is granted by the Japan Society for the Promotion of Science (JSPS) through the "Funding Program for World-Leading Innovative R&D on Science and Technology (FIRST Program)," initiated by the Council for Science and Technology Policy (CSTP) and in part supported by KAKENHI (Grant No. 23104702) from MEXT, Japan (T. M.).

**FIGURES**

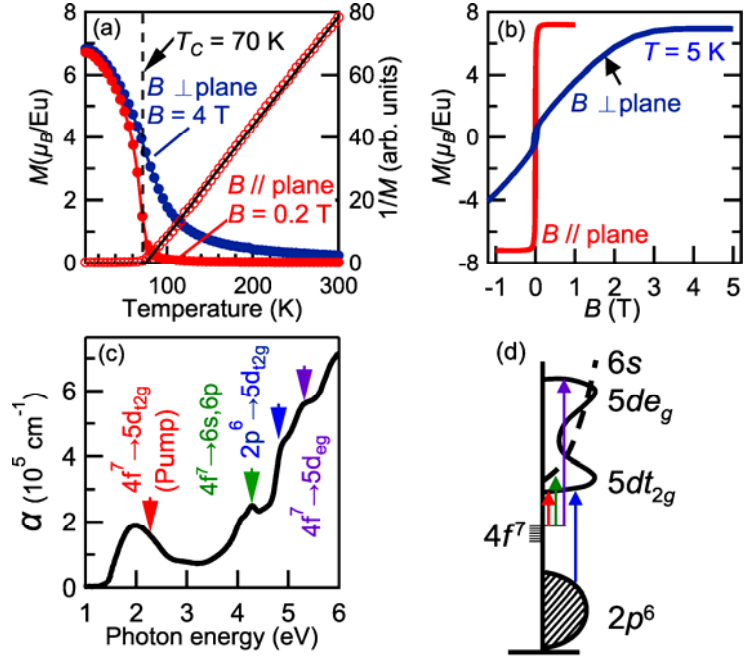

FIG. 1 (color online). (a), Temperature dependences of magnetization and inverse values of magnetization in EuO thin films under in-plane [red (medium gray)] and out-of-plane [blue (dark gray)] configurations at magnetic fields of $B = 0.2$ T and 4 T, respectively. (b), Static magnetization-versus-magnetic field curves taken under the configurations of out-of-plane and of in-plane at $T = 5$ K. (c), optical absorption spectrum in a EuO thin film at $T = 300$ K. Assignments are also depicted. (d), Schematic diagram of the density of states in EuO.

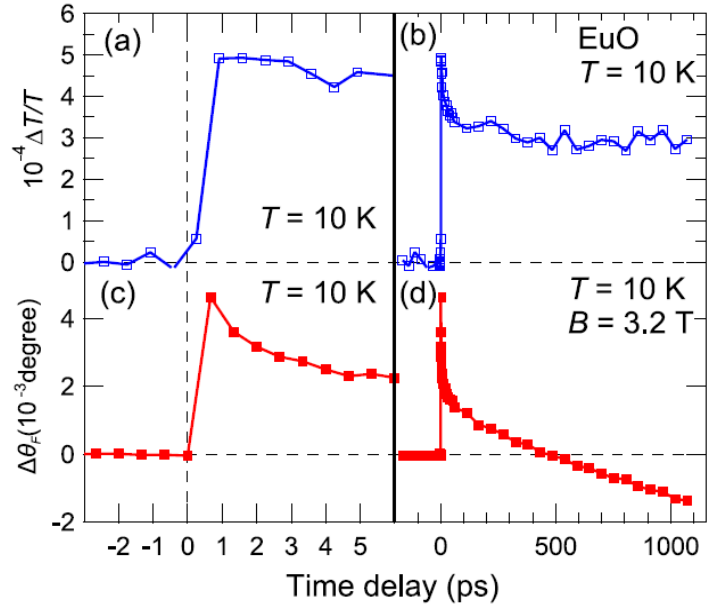

FIG. 2 (color online). Time evolution of $\Delta T/T$ (a),(b) and $\Delta\theta_F$ at 1.9 eV at 10 K (c),(d).

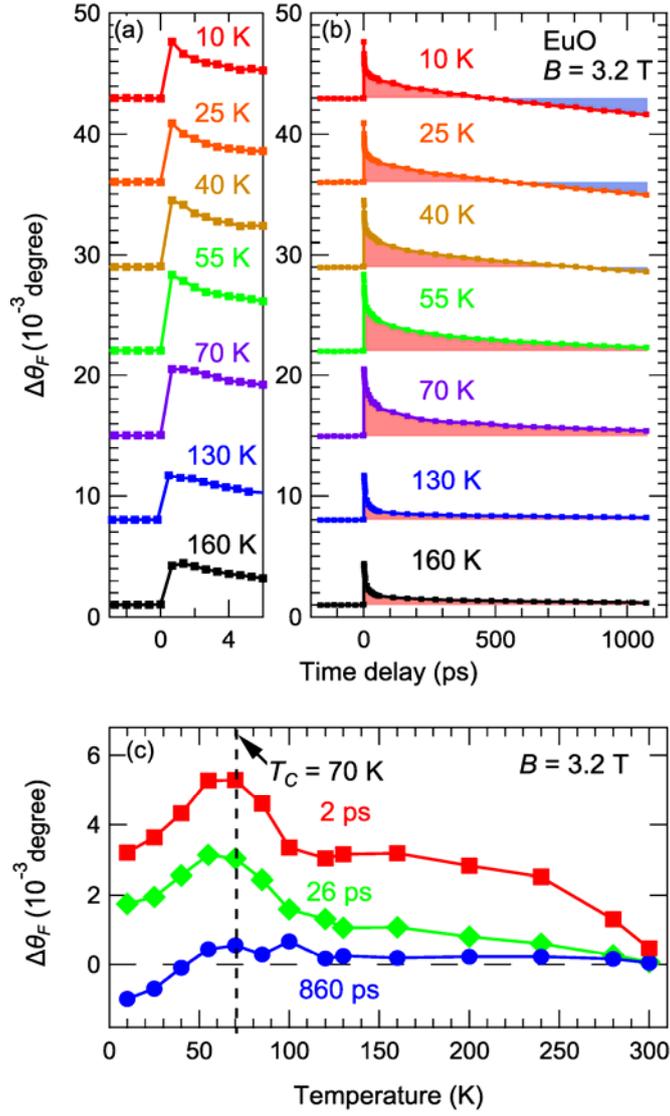

FIG. 3 (color online). Temporal traces of $\Delta\theta_F$ at different temperatures under out-of-plane 3.2-T magnetic field [(a),(b)]. Right (a) and left (b) panels respectively depict the fast and slow temporal regions. All traces are offset for clarity. (c) Temperature dependence of magnetization changes at different time delays: 2 ps (squares), 26 ps (diamonds) and 860 ps (circles).

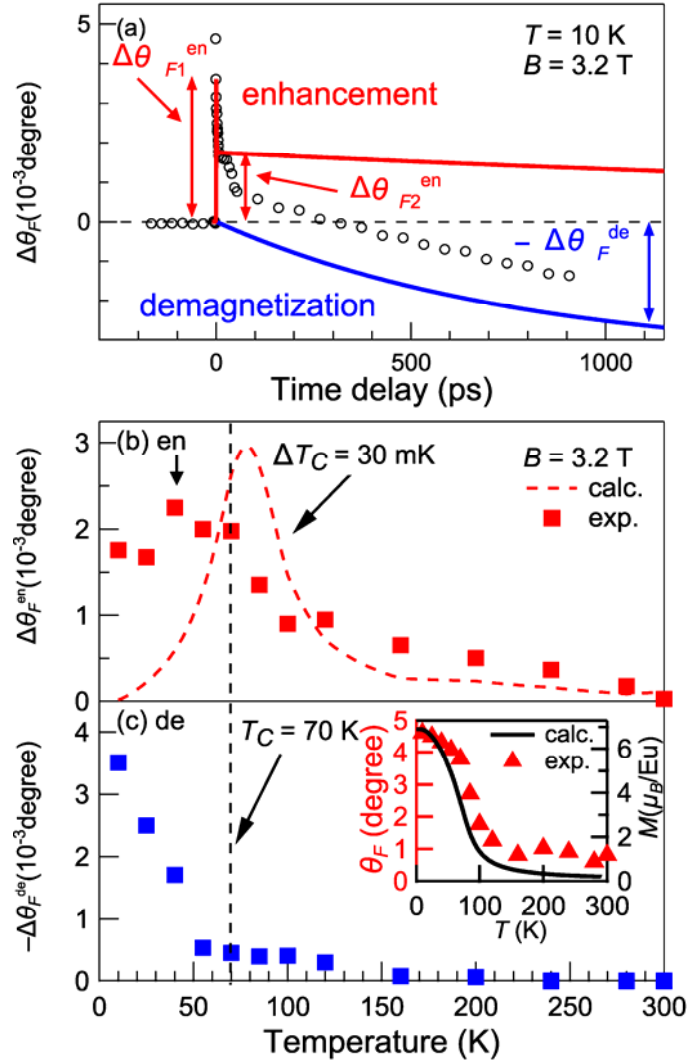

FIG. 4 (Color online): (a), Deconvolution of the photoinduced Faraday rotation change $\Delta\theta_F$ into enhanced magnetization $\Delta\theta^{en}_{F1(2)}$ [red (medium gray) curves] and demagnetization $\Delta\theta^{de}_F$ [a blue (dark gray) curve] components at $T = 10$ K and $B = 3.2$ T. (b), Temperature dependences of enhanced magnetization $\Delta\theta^{en}_{F2}$ (red closed squares) components. The calculation for the effects of $\Delta T_C$ on $\Delta M\theta_F/M$ (a dashed line for $\Delta T_C = 30$ mK) is also shown. (c), Temperature dependence of demagnetization $\Delta\theta^{de}_F$ (blue squares) component. The inset shows the static Faraday rotation ($\theta_F$) versus temperature ($T$) at $B = 3.2$ T (closed red squares) along with calculated magnetization (solid black curve).